\documentclass{aa}           

\usepackage[normalem]{ulem}
\usepackage{graphicx}

\usepackage{txfonts}

\usepackage{amsfonts}
\usepackage{amsmath}
\usepackage{amssymb}

\usepackage{subfigure}  

\usepackage{longtable}
\usepackage{multirow}
\usepackage{tabularx}              
\usepackage{array}     

\usepackage{hyperref}

\begin{document} 

\title{VLTI/PIONIER images the Achernar disk swell\thanks{Based on observations performed at
    ESO, Chile under VLTI/PIONIER program IDs 087.D-0150,
    189.C-0644, and 093.D-0571.\protect\\
    \email{gaetan.dallavedova@gmail.com} \protect\\ 
    \email{florentin.millour@oca.eu}}
  }
\authorrunning{G. Dalla Vedova et al.}
\titlerunning{Disk-formation around Achernar}

\author{G. Dalla Vedova\inst{1}
\and F. Millour\inst{1}
\and A. Domiciano de Souza\inst{1}
\and R. G. Petrov\inst{1}
\and D. Moser Faes\inst{1,5}
\newline A. C. Carciofi\inst{5}
\and P. Kervella\inst{2,3}
\and T. Rivinius  \inst{4}
}

\institute{{Universit\'e C\^ote d'Azur, Observatoire de la C\^ote
    d'Azur, CNRS, Lagrange, Blvd de l'Observatoire, CS 34229, 06304
    Nice cedex 4, France}
\and{Unidad Mixta Internacional
    Franco-Chilena de Astronom\'{i}a (UMI 3386), CNRS/INSU, France \&
    Departamento de Astronom\'{i}a, Universidad de Chile, Camino El
    Observatorio 1515, Las Condes, Santiago, Chile}  \and{LESIA (UMR
    8109), Observatoire de Paris, PSL, CNRS, UPMC,
    Univ. Paris-Diderot, 5 place Jules Janssen, 92195 Meudon, France}
\and{ESO --- European Organisation for Astronomical Research in the
    Southern Hemisphere, Santiago, Chile, Casilla 19001}  
\and{Instituto de Astronomia, Geof{\'\i}sica e Ci{\^e}ncias
    Atmosf{\'e}ricas, Universidade de S{\~a}o Paulo (USP), Rua do
    Mat{\~a}o 1226, Cidade Universit{\'a}ria, S{\~a}o Paulo, SP -
    05508-900, Brazil}}
\date{Received; Accepted}

\abstract
{The mechanism of disk formation around fast-rotating Be stars is not
  well understood. In particular, it is not clear which mechanisms operate, in addition to fast rotation, to produce the observed variable ejection of matter. The star Achernar is a privileged laboratory to probe these additional mechanisms because it is close, presents B$\rightleftharpoons$Be phase variations on timescales ranging from $\sim 6$ yr to $\sim 15$ yr, a companion star was discovered around it, and probably presents a polar wind or jet.}
{Despite all these previous studies, the disk around Achernar was never directly imaged. Therefore we seek to produce an image of the photosphere and close environment of the star.}
{We used infrared long-baseline interferometry with the PIONIER instrument at the Very Large Telescope Interferometer (VLTI) to produce reconstructed images of the photosphere and close environment of the star over four years of observations. To study the disk formation, we compared the observations and reconstructed images to previously computed models of both the stellar photosphere alone (normal B phase) and the star presenting a circumstellar disk (Be phase).}
{The observations taken in 2011 and 2012, during the quiescent
  phase of Achernar, do not exhibit a disk at the detection limit of the
  instrument. In 2014, on the other hand, a disk was
  already formed and our reconstructed image reveals an extended H-band continuum excess flux. Our results from interferometric imaging are also supported by several H$\alpha$ line profiles showing that Achernar started an emission-line phase sometime in the beginning of 2013. The analysis of our reconstructed images shows that the 2014 near-IR flux extends to $\sim 1.7-2.3$ equatorial radii. Our model-independent size estimation of the H-band continuum contribution is compatible with the presence of a circumstellar disk, which is in good agreement with predictions from Be-disk models.}
{}

\keywords{stars, individual: Achernar -- stars: rotation -- stars:
  imaging -- (stars:) circumstellar matter -- instrumentation:
  interferometers -- techniques: high angular resolution} 
\maketitle

\section{Introduction} \label{sec:introduction}

With a spectral type B3Vpe, a visual magnitude of $m_v=0.46$, and a distance of $42.75$\,pc
\citep{2007AandA...474..653V}, Achernar ($\alpha$ Eridani, HD 10144)
is the brightest and the nearest Be star in the sky. A Be star is a non-supergiant B-type star with sporadic episodes of Balmer lines emission,  attributed to the circumstellar environment.

Achernar is a fast-rotating star, which is spinning at 88\% of its critical
velocity and has a projected rotation velocity at the stellar surface
$v\sin i = 220-270$ km\,s$^{-1}$
\citep{2006AandA...446..643V, 2014AandA...569A..10D}. Fast rotation flattens the stellar
globe and makes the equator cooler than the poles with the so-called gravity darkening or
von Zeipel effect \citep{1924MNRAS..84..684V}.
This star was extensively observed in the past, and there is a transient
circumstellar disk that appears and disappears in a possible cyclic way. \citep{2006AandA...446..643V} argued that this cycle has a period of $\sim 14-15$ years. 
However, a more recent work \citep{Faes2015_PhD} suggests a shorter timescale of variability that is not necessarily periodic. Achernar started an outburst in the beginning of 2013, after seven years of quiescence, as reported by \citep{2015IAUS..307..261F}.

Achernar also hosts a companion star at a
projected separation going from 50 to 300\,milli-arcseconds
\citep[noted hereafter as mas, i.e., from 2 to
13\,au;][]{2007AandA...474L..49K,2008AandA...484L..13K}. The orbit
remains to be determined, but preliminary results indicate a period of
a few years and a mass of the secondary of between 1 and 2 solar masses
(Kervella, Domiciano de Souza et al. in prep.).

\citet{2014AandA...569A..10D} proposed a model of the photosphere, the Roche-von Zeipel model 
(hereafter referred to as the RvZ model) in which the star is deformed by fast rotation and its photospheric
temperature changes with latitude. In that paper, the authors used
the previously determined values of the equatorial angular diameter to 1.99\,mas, orientation angle of the polar axis to 36.9\,degrees, and flattening ratio to 1.29. These authors performed an image reconstruction of
Achernar in 2011, in which no signatures of an additional
circumstellar component within a $\sim\pm1\%$ level of intensity could
be detected.

In this work, we compare the appearance of Achernar at four epochs
from 2011 to 2014, using state-of-the-art models and reconstructed
images from the PIONIER instrument. We show, for the first time,
images of the disk forming around a Be star.

The paper is organized as follows: 
In Sec.~\ref{sec:observationsdata}, we introduce the campaign of observation and  data collected.
In Sec.~\ref{sec:imaging}, we present the image reconstruction process
that we used to reconstruct the images of the object.
In Sec.~\ref{sec:modelling}, we compare the data with the models.
In Sec.~\ref{sec:imgmodel}, we compare the images of both object and
model showing evidence of the forming circumstellar disk around
Achernar.
Finally, in Sec.~\ref{sec:discussion}, we discuss these results in the context of the current overview of the system.

\section{Observations and data} \label{sec:observationsdata} 

Several campaigns of observations of Achernar have been carried out
between 2011 and 2014 with the PIONIER interferometer at the Very Large Telescope Interferometer (VLTI)
\citep{2003ESASP.522E...5G,2011AandA...535A..67L}. This instrument combines the
light beams from four telescopes and measures the light spatial coherence
(related to the shape of an object) through six simultaneous visibilities
and four closure phases in the H band (1.65\,$\mu$m). The data have been
reduced using the standard procedure \textit{pndrs}, especially
developed for the PIONIER instrument \citep{2011AandA...535A..67L}. The log of observations is
presented in Tab.~\ref{tab:LogData}. We define four different epochs of observations in our analysis: August/September 2011, September 2012, September 2013, and September 2014.

\begin{table}
\caption{Log of the PIONIER observations (four simultaneous
  telescopes, except in 2013 when there were 3 telescopes). We defined 4 datasets relative to different epochs: August/September 2011, September 2012, September 2013, and September 2014.}
\label{tab:LogData}
\centering
\begin{tabular}{lcccc}
\hline \hline
Date&Nb obs.&Nb $\lambda$ & Config. & Epoch\\
\hline\hline
2011 Aug. 06 & 4  & 3 & A1-G1-K0-I1 & 2011\\
2011 Sep. 22 & 10 & 7 & A1-G1-K0-I1 & 2011\\
2011 Sep. 23 & 9  & 7 & A1-G1-K0-I1 & 2011\\
\hline
2012 Sep. 16 & 9  & 3 & A1-G1-K0-I1 & 2012\\
2012 Sep. 17 & 3  & 3 & A1-G1-K0-I1 & 2012\\
\hline
2013 Sep. 02 & 2  & 3 & G1-K0-J3    & 2013\\
2013 Sep. 04 & 1  & 3 & G1-K0-J3    & 2013 \\
\hline
2014 Sep. 21 & 5  & 3 & A1-G1-K0-J3 & 2014\\
\hline
\end{tabular}
\end{table}

\begin{figure*}
\centering
\hspace{-.5cm}\includegraphics[width=0.70\textwidth]{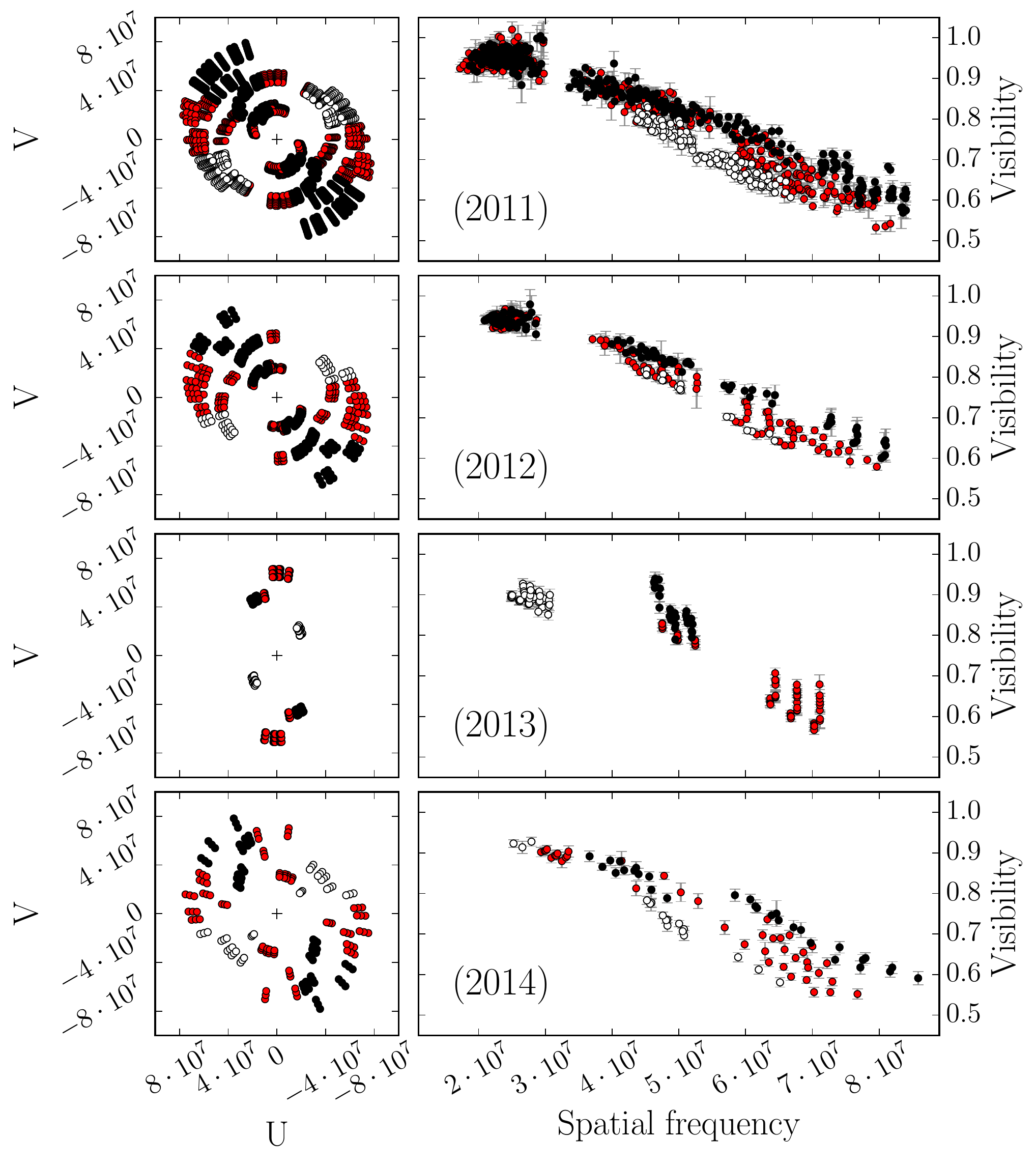}
\caption{{\bf Left:} spatial frequencies (u,v) coverage for each dataset. {\bf Right:} visibility
  amplitudes (with their respective error bars) as a function of
  spatial frequency ($B/\lambda$ in units of rad${^{-1}}$) for each dataset. Epochs
  2011, 2012, 2013, and 2014 are represented from top to
  bottom. 
  Colors indicate position angle (PA) ranges. Black circles
  correspond to regions around polar axis of the star $PA_\mathrm{pol}=36.9^{\circ}$
  ($14.4^{\circ}\leq PA \leq 59.4^{\circ}$), hollow circles match the
  equatorial axis $PA_\mathrm{eq}=126.9^{\circ}$ ($104.4^{\circ}\leq PA \leq
  159.4^{\circ}$), and red circles correspond to the other directions.}
\label{fig:data}
\end{figure*}

Fig.~\ref{fig:data} presents (u,v) coverage and visibility as a function of spatial frequency for each dataset. 
The slope of the visibility function is different in
the polar and equatorial directions owing to the flattening of
the photosphere of the star (see Fig.~\ref{fig:data}). 
We do not show the closure phases as they are almost all compatible with zero.

The (u,v) coverage for each dataset, except that of 2013, is
very uniform, which is a relevant condition for image
reconstruction. The angular resolution of the observations is sufficient
to reconstruct the global shape of the object and especially resolve
any extended structure around the star.

\section{Image reconstruction procedure} \label{sec:imaging}

We used the \texttt{MiRA} software developed by
\citet{2008SPIE.7013E..43T}\footnote{\texttt{MiRA} is available
    from \protect\url{https://cral.univ-lyon1.fr/labo/perso/eric.thiebaut/?Software/MiRA}},
combined with a \textit{Monte Carlo} (MC) approach to reconstruct the
images for the observations in 2011, 2012, and 2014 because the (u,v) coverage
in 2013 is insufficient.
The \texttt{MiRA} software is based on an iterative process, starting from an
initial image, modifying it to find the pixel values $x_i$ which
minimize a distance between the interferometric data $d_i$ and the
same interferometric observables calculated from the image $m_i$.
This software uses a Bayesian framework and an advanced gradient descent
algorithm to find the best solution to this ill-posed problem.
The distance between the interferometric data and the image is posed
as $\chi^2 = \sum_i\frac{(d_i-m_i)^2}{\sigma_i^2}$, where $\sigma_i$
is the data noise.
The process of minimizing this distance is conducted under additional
constraints of positivity, normalization, and generic constraints on
the image, called regularization. 
The regularization has a global weight relative to the distance
between data and image. This weight is usually called
hyperparameter and noted as $\mu$. The minimization criterion for the
image reconstruction process is then $C = \chi^2 + \mu \times
R$. Finding the right value for $\mu$ is usually difficult, but new
ideas using MC methods provide a proxy to this value.

We
used a Gaussian prior with varying sizes to concentrate the flux in the central part of the image (see next paragraph). We set
the pixel size of the image reconstruction to 0.07\,mas.
In our case, we used 1000 reconstructed \texttt{MiRA} images of
Achernar for  both tuning the hyperparameter, with the L-curve method 
\citep{2014AandA...564A..80K} following
\citet{DallaVedova2015_PhD,2014AandA...569A..10D}, 
and for producing the final image using the Brute-Force Monte Carlo
technique \citep[BFMC; see][]{2012SPIE.8445E..1BM}.

The L-curve method is a graph of $\mu$ as a function of the distance
to the data $\chi^2$. The obtained curve shows a plateau of low
$\chi^2$ values for lower values of $\mu$, and a steep increase
afterward. Selecting the reconstructions among these low $\chi^2$
values plateau ensures a good match of the reconstruction to the data,
while preserving the benefits of regularization.

The BFMC method works as follows. We reconstructed 1000 images with
random values of $\mu$ (in the $[1 : 1 \cdot 10^9]$ range), a random
start image, and random size of the Gaussian prior, scanning through
the parameter space of image reconstruction. We kept the subset of
reconstructed images with the smaller distance to the data, in our
case $\chi^2 \leq 2 \chi^2_{\rm min}$, and the optimum regularization
strength evaluated with the L-curve method, in our case $200 \leq \mu
\leq 300$. 
We then calculated the average image by calculating the mean of each
pixels based on the selected reconstructed images, after centering
them. 
Centering the images is essential to saving the angular resolution of the observations because
closure phases do not contain the absolute astrometric position of the
star \citep[][section 11.7.2]{van2012astrometry} and the individual
reconstructed images end up in a random position. The companion star is likely outside the $\approx$200\,mas field of
view of the observations and does not affect the image reconstruction.

\begin{figure*}[htbp]
\includegraphics[width=.95\textwidth]{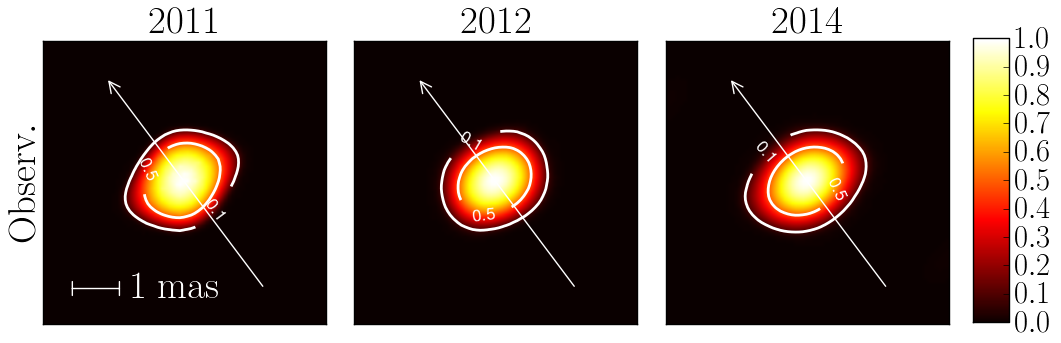}
\caption{Reconstructed image of Achernar for, from  left to right, the 2011, 2012, and
  2014 datasets. East is left and north is up. The long arrow shows the
  previously estimated position angle of the rotation axis of the star on
  the plane of sky. The dynamic of the images is equal to 1.}
\label{fig:images}
\end{figure*}

The Fig.~\ref{fig:images} presents the resulting
reconstructed images of Achernar for the 2011, 2012, and 2014 sets of observations. The 2011 and 2012 images show a flattened star similar to previous reconstructions \citep[see][]{2014AandA...569A..10D}, but the 2014
image show extended equatorial emission in addition to the flattened
photosphere, which we need to compare with the noise and image reconstruction
artefacts level to determine if this is a real structure or not.

\begin{figure*}[htbp]
\includegraphics[width=.981\textwidth]{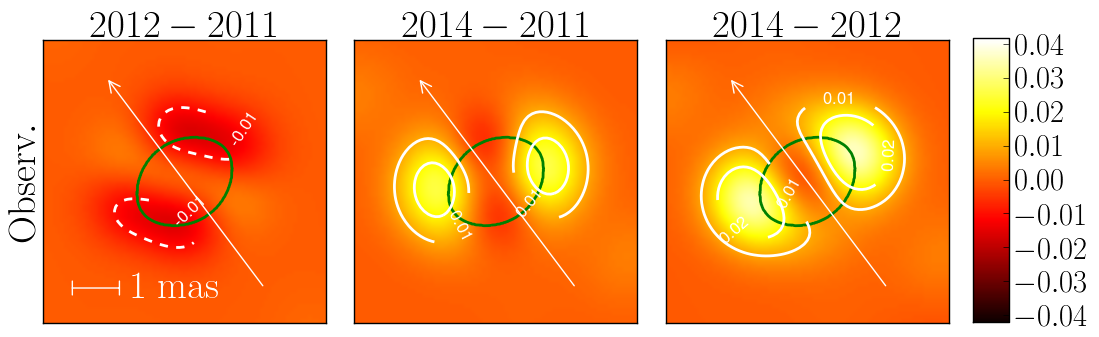}
\caption{Subtraction of the Achernar reconstructed between different epochs. From left to
  right, subtraction between the 2012 and 2011 images, subtraction between the 2014 and 2011 images, and subtraction
  between the 2014 and 2011 images.
  East is left and north is up. The long arrow shows the
  previously estimated position angle of the rotation axis of the star on
  the plane of sky.
  The green ellipse indicates the delimiting profile of the RvZ model of the
  photosphere model of Achernar, and the images are highlighted
  with contours  -0.02, -0.01, 0.01, and 0.02.}
\label{fig:imagesSous}
\end{figure*}
A way to investigate if this extended equatorial structure is real is to compare the images at different epochs. Fig.~\ref{fig:imagesSous} shows
the subtraction between the reconstructed images of Achernar over the three epochs: 2012-2011, 2014-2011, and 2014-2012.
The two subtracted images, including the 2014 observations (2014-2011 and 2014-2012), present an excess of flux of $\sim 4\%$ that is not present in the 2011-2012 subtracted image. This excess can be attributed to a gas disk forming around the main star.

These subtracted images allow us to estimate the size and orientation of the close circumstellar disk independently from any model. 
From the subtracted images 2014-2011 and 2014-2012, we estimate the equatorial disk diameter in the H band to be between 3.4(4.2) and 3.7(4.5)\,mas, considering the external limits of the 0.02(0.01) level contours. This corresponds to the angular diameter of the circumstellar region emitting most of the H-band flux. The extreme values of these angular size estimates translates into a disk radius between 1.7 and 2.3 stellar equatorial radii (adopting the equatorial angular diameter of 1.99~mas).

This model-independent size estimate of the H-band emission region is in good agreement with predictions from theoretical models. Indeed, the formation loci of continuum emission at various wavelengths for different Be disk models were computed by \citet{Rivinius2013_v21p69, Vieira2015_v454p2107-2119, 2011IAUS..272..325C}. According to the results of these authors, most ($>80\%$) of the H-band disk flux is emitted within a radius between $\sim1.5$ to $\sim4$ times the photospheric equatorial radius, which matches well our estimated disk size.

\section{Modeling} \label{sec:modelling}

We considered two models of Achernar to check
if there are indeed variations in the shape of the star between all
the considered epochs.

\subsection{Roche-von Zeipel model with CHARRON} \label{sec:charron}

\begin{figure}
\centering
\includegraphics[width=.5\textwidth]{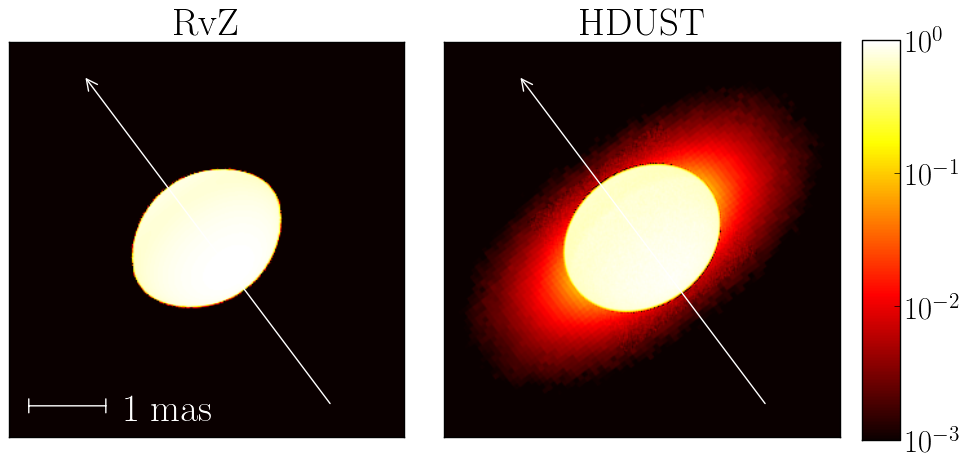}
\caption{Logarithmic of the normalized intensity maps in the H band for the two models used in this paper. {\bf Left:} RvZ model computed with the CHARRON code with the model parameter values given by \citet{2014AandA...569A..10D}. {\bf Right: } HDUST model with an ellipsoidal photosphere similar to the RvZ model and a circumstellar viscous decretion disk model \citep[further details in][Chapter 6]{Faes2015_PhD}. The flux ratio of photosphere and disk model relative to the purely photospheric emission is 1.023.}
\label{fig:model_images}
\end{figure}

As a starting point, we adopted the same photospheric model as in
\citet{2014AandA...569A..10D}. We adopted the Roche approximation (rigid rotation and mass
concentrated in the stellar center), which is well adapted for rapidly
rotating stars, to model the photosphere of
Achernar. We use the code CHARRON (Code for High Angular
Resolution of Rotating Objects in Nature) described in detail by
\citet{Domiciano-de-Souza2012_v545pA130,
  Domiciano-de-Souza2012_vp321-324,
  Domiciano-de-Souza2002_v393p345-357}. 
The star is flattened (equatorial radius larger than the polar radius)
because of the important centrifugal force generated by the high
rotation velocity of the star. 
The effective temperature $T_\mathrm{eff}$ at the surface
depends on the colatitude $\theta$ due to the decreasing effective
gravity $g_\mathrm{eff}$ (gravitation plus centrifugal acceleration)
from the poles to the equator (gravity darkening effect). 
The resulting map can be seen in the left panel of Fig.~\ref{fig:model_images}. This map corresponds to the model parameter values determined by
\citet{2014AandA...569A..10D}.

From this model and the (u,v) coordinates of the observations, we
calculated the corresponding synthetic visibilities and computed the
residuals between the model and observations at all epochs.

\begin{figure*}[htbp]
\centering
\includegraphics[width=0.48\textwidth]{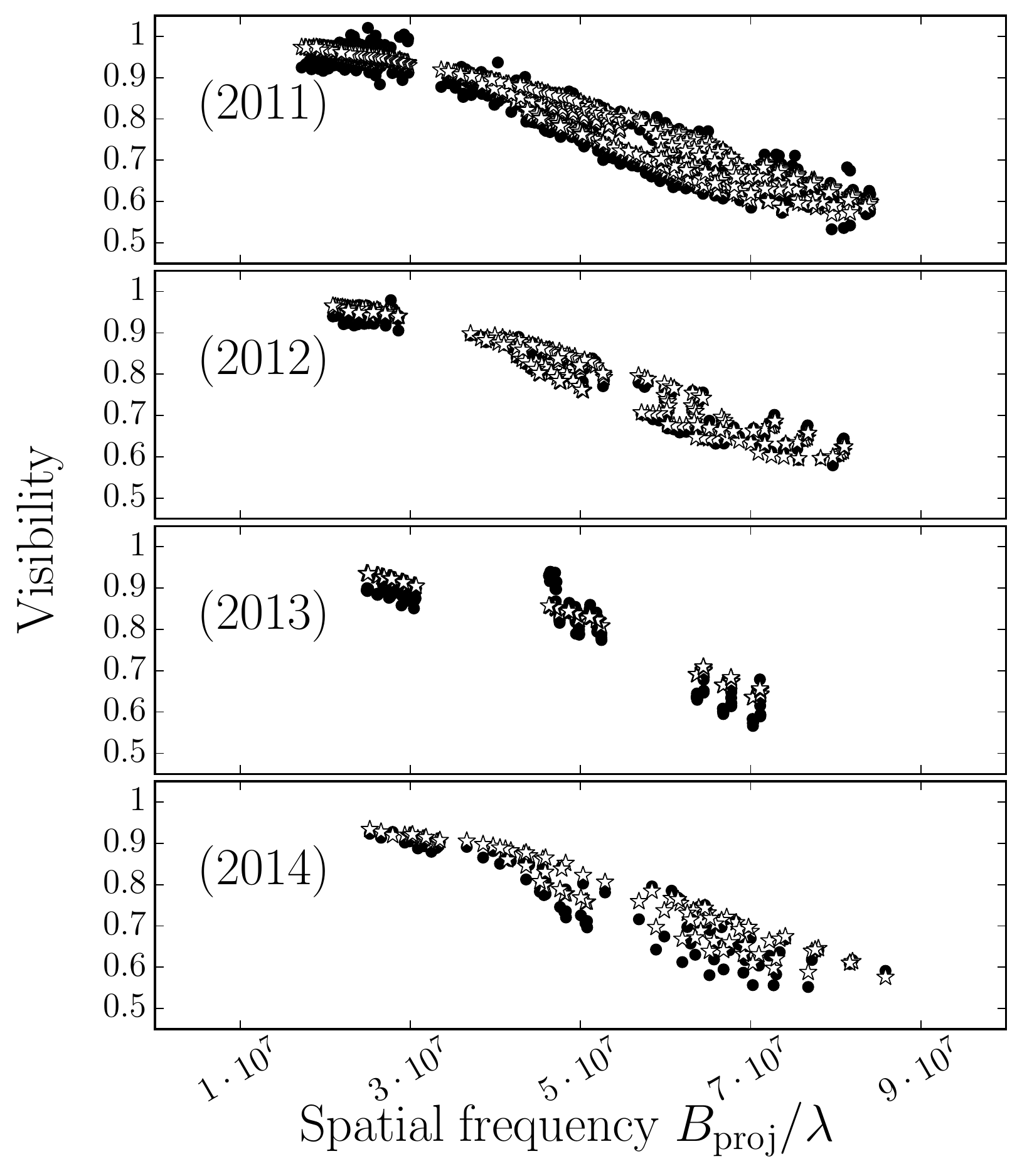}
\includegraphics[width=0.48\textwidth]{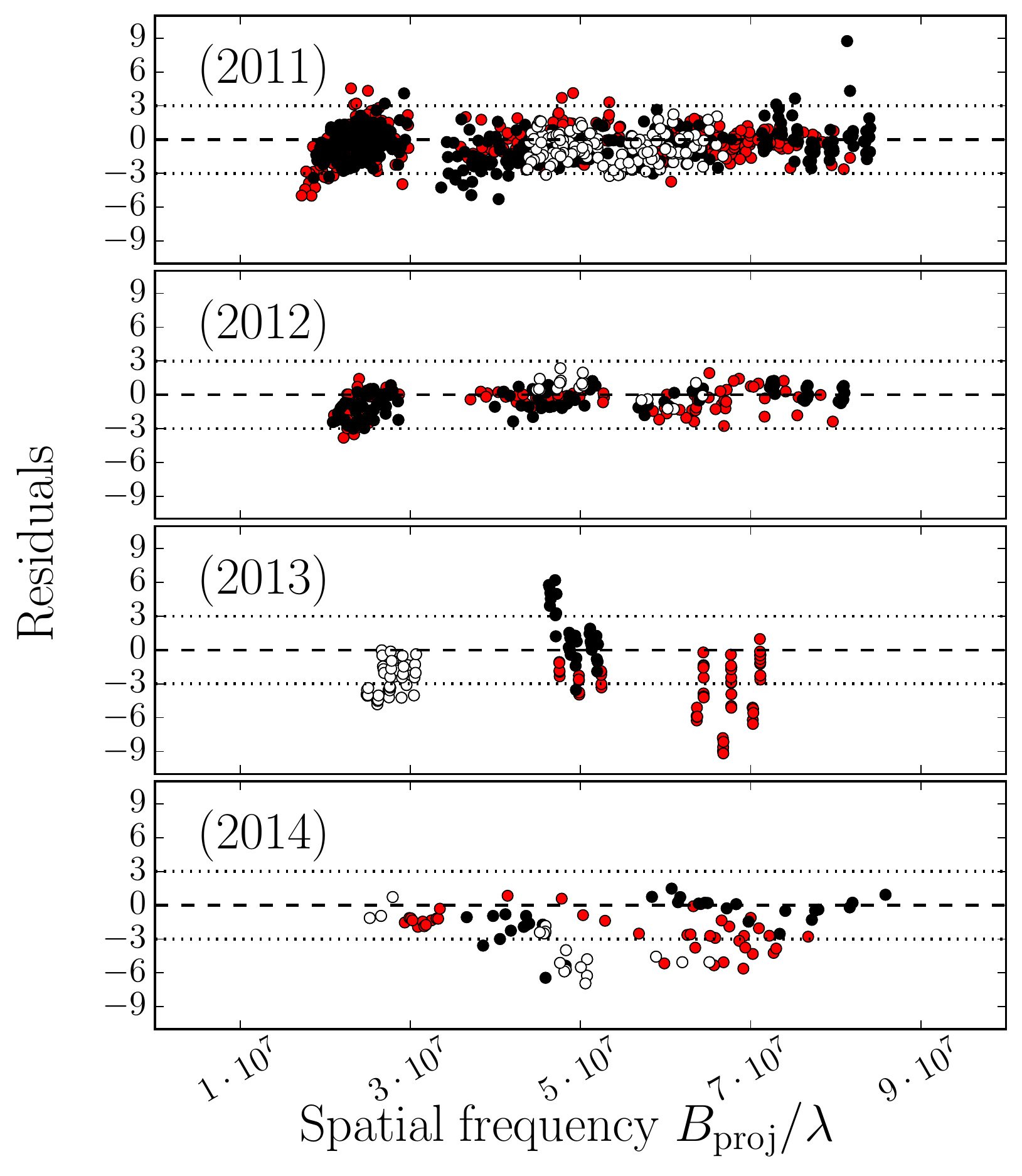}
\caption{Roche-von Zeipel model compared to the Achernar data. {\bf Left:} Observed
  (black dots) and modeled (white stars) visibilities of
  Achernar. {\bf Right:} Residual of the visibility between the
  observations and the RvZ model as a function of the spatial
  frequency for each dataset. Colors indicate polar and equatorial
  axes, respectively, as in Fig. \ref{fig:data}.}
\label{fig:residual}
\includegraphics[width=0.48\textwidth]{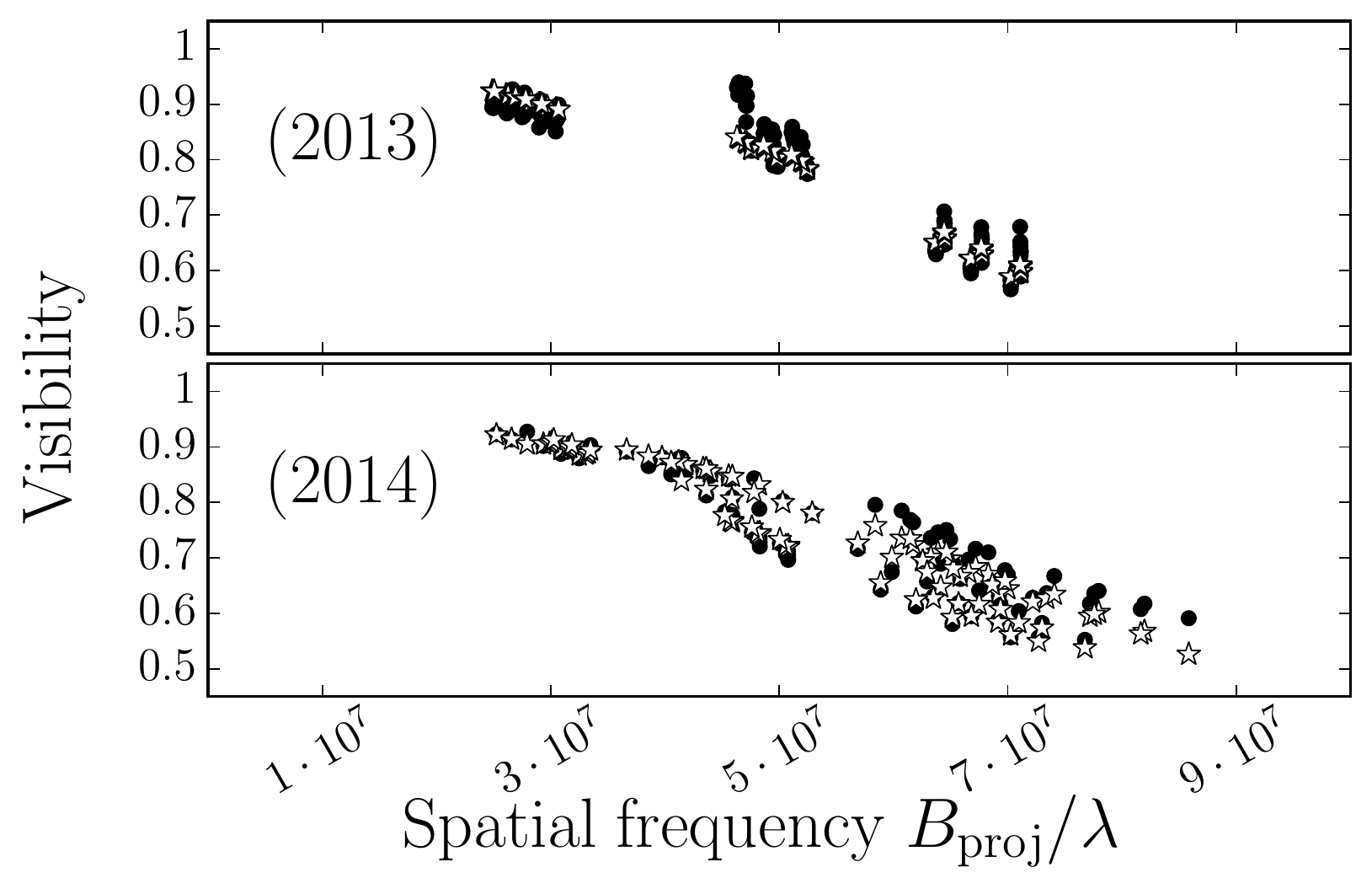}
\includegraphics[width=0.48\textwidth]{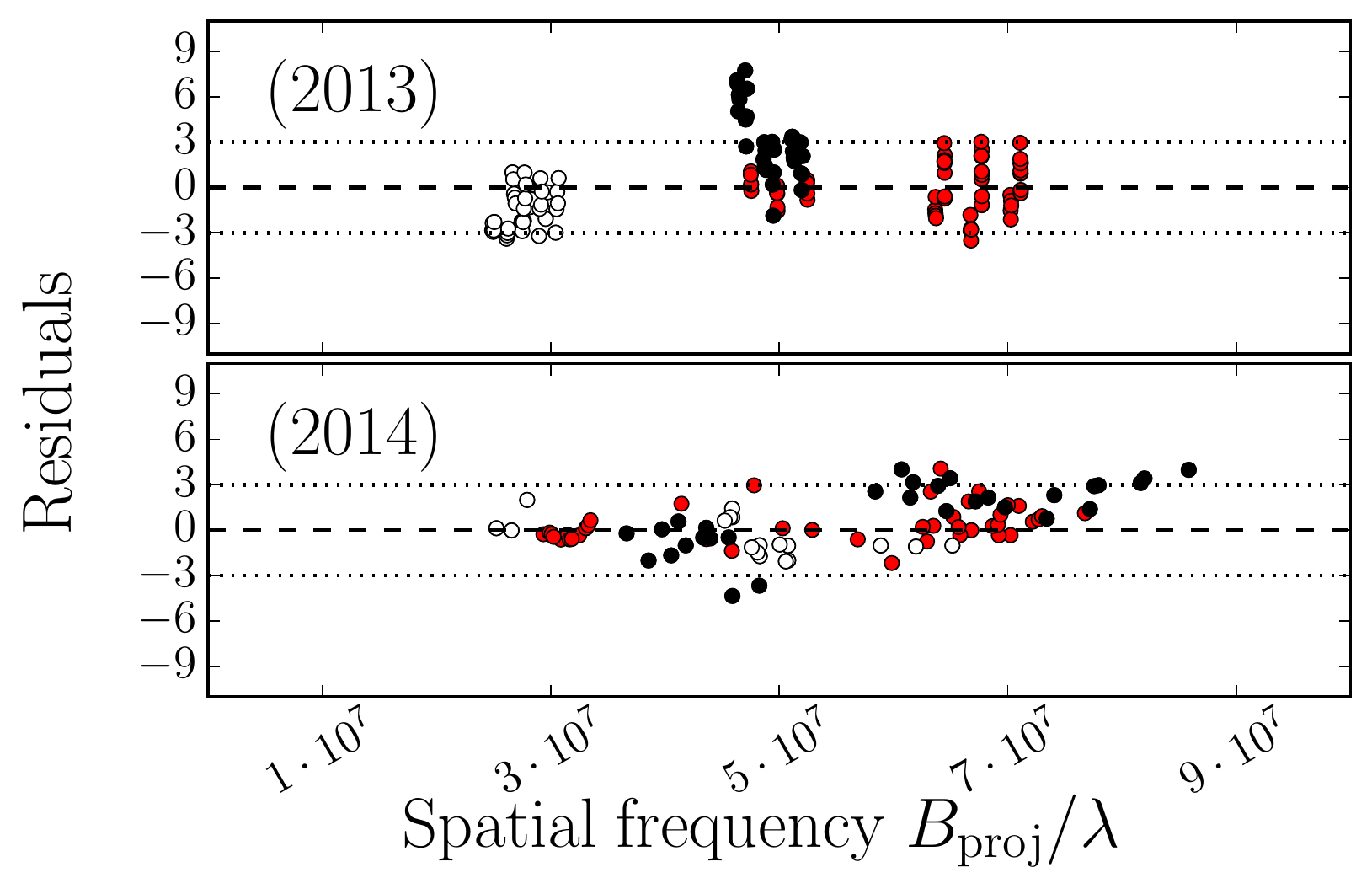}
\caption{Visibilities of HDUST model compared to the Achernar
    visibilities. {\bf Left:} Observed (black dots) and modeled (white stars)
    visibilities of Achernar. {\bf Right:} Residual of the visibility
    between the observations and the HDUST model as a function of the
    spatial frequency for each dataset. Black and white dots indicate
    polar and equatorial axes, respectively, such as in
    Fig. \ref{fig:data}.}
\label{fig:residual_HDUST}
\end{figure*}
Fig.~\ref{fig:residual} shows a comparison of these modeled and
observed visibilities along with the computed residuals as a function
of the spatial frequencies for each dataset. The model visibilities
match the data in 2011 and 2012, but we can see clear departures in
2013 (especially around the frequency $3\times10^7$ rad$^{-1}$) and
2014 (especially between $4\times10^7$ rad$^{-1}$ and $7\times10^7$
rad$^{-1}$).

This is confirmed by the residuals (right panel of Fig.~\ref{fig:residual}), which are contained between the $\pm3\sigma$
typical limit, with a few outliers in 2011 and 2012, but that clearly diverge above $3\sigma$ in 2013 and 2014, especially in the
equatorial direction (hollow dots). This excess residual provides
evidence of the presence of an elongated structure around the stellar
equator, confirming the excess equatorial emission seen in the
reconstructed images.

\begin{figure*}
\centering
\includegraphics[width=.981\textwidth]{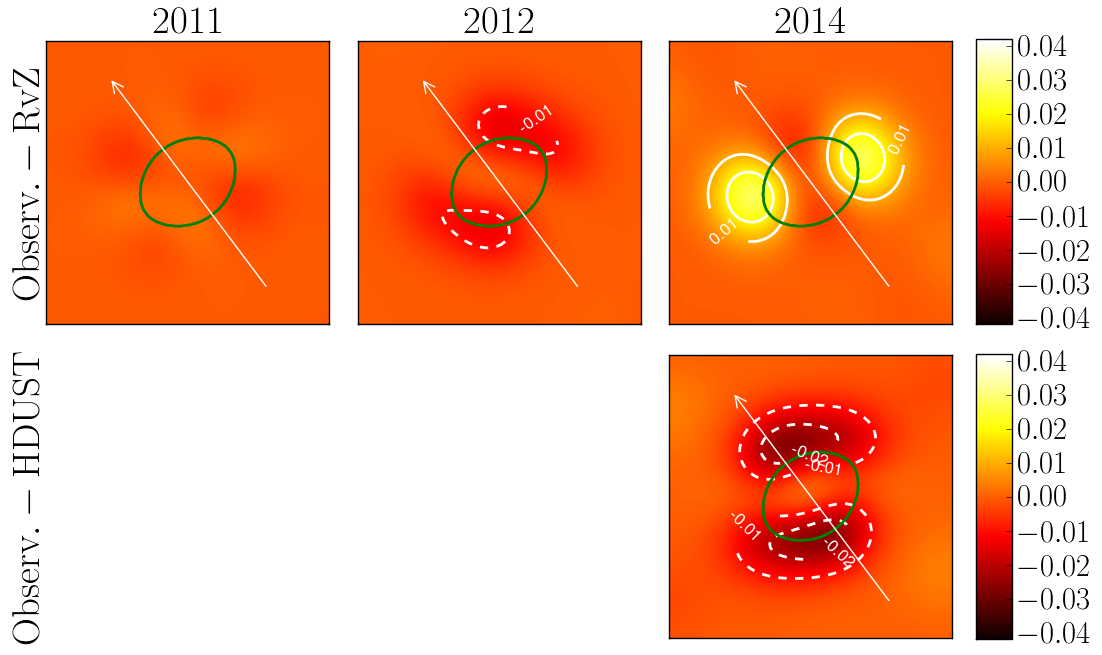}
\caption{Subtraction between the reconstructed image of Achernar and both the RvZ
  and HDUST models.
  On the \textbf{top row},
  subtraction between the images reconstructed from the observed
  data and artificial data calculated from the RvZ model of the
  photosphere of the star. 
  On the \textbf{bottom row}, subtraction
  between the reconstructed image from the observed data and from the
  artificial data calculated form the HDUST model of the photosphere
  and circumstellar disk. 
  The dynamic of the images is equal to 1.   
  The green ellipse indicates the profile of the RvZ model of the
  photosphere model of Achernar. The subtraction images are highlighted
  with contours -0.02, -0.01, 0.01, 0.02.
  East is left and north is up. The long arrow shows the
  previously estimated position angle of the rotation axis of the star on
  the plane of sky.
  All datasets have been submitted to the same image
  reconstruction procedure.}
\label{fig:imagesSous2}
\end{figure*}

\subsection{Circumstellar disk model with HDUST} \label{sec:hdust}

To model this discrepancy between the RvZ model and observed data
in 2013 and 2014, we add a circumstellar environment to Achernar by
using the 3D, NLTE Monte Carlo radiative transfer code HDUST, which was
developed by \citet{Carciofi2006_v639p1081-1094}. We briefly describe
HDUST here, while a more detailed description is provided in the above
paper. The central star is modeled by an oblate ellipsoid of
revolution with identical flattening as the Roche model. Gravity
darkening is also included in the model to obtain the surface
distribution of effective temperature.

The circumstellar environment is modeled by iterating on the
radiative and statistical equilibrium equations to calculate
the H-level populations, electron temperatures, and hydrostatic
equilibrium density for all grid cells. These calculations include
collisional and radiative processes (bound-bound, bound-free, and
free-free). In agreement with several recent results on Be star disks,
the adopted physical structure and velocity field of the circumstellar
disk of Achernar follow the viscous decretion disk (VDD) model
\citep[e.g.,][]{Rivinius2013_v21p69, 2011IAUS..272..325C}. Using the oblate photosphere and VDD prescription, \citet[][Chapter 6]{Faes2015_PhD} computed a HDUST disk model that roughly agrees with the polarization and H$\alpha$ spectra of Achernar in 2014. The values for the equatorial radius, flattening, and gravity darkening coefficient adopted for the oblate ellipsoid photospheric model are those from \citet{2014AandA...569A..10D}. 
The resulting intensity map can be seen in
Fig.~\ref{fig:model_images}, right, and the corresponding H$\alpha$
line profile is given in Fig.~\ref{fig:spectra}. 

Following our method in Sect.~\ref{sec:charron}, we computed visibilities out of this
model map and compared them with the observed visibilities in
Fig.~\ref{fig:residual_HDUST}. There is no point in comparing the HDUST
model with 2011 and 2012, where the RvZ model neatly matches the data
\citep[a comparison that was already carried out
in][]{2014AandA...569A..10D}, therefore, Fig.~\ref{fig:residual_HDUST}
only shows 2013 and 2014. 
We see a clear improvement of the visual match of visibilities between
the RvZ model and HDUST model. This is confirmed 
by the residual plot, where the residuals shrink within
$\pm3\sigma$, and the computed $\chi^2$ values from
Table~\ref{tab:Chi2_compar} go from 9.7 to 4.8, and 7.9 to 4.4 in
2013 and 2014, respectively.

\begin{table}
\caption{Computed reduced $\chi^2$ values for the different models
  considered in this paper. The reduced $\chi^2$ for the HDUST model
  in 2011 \& 2012, with values around 5, is not relevant for this
  study.}
\label{tab:Chi2_compar}
\centering
\begin{tabular}{lcccc}
\hline \hline
Date& Red. $\chi^2$ CHARRON & Red. $\chi^2$ HDUST\\
\hline\hline
2011 & 1.8 & - \\
2012 & 1.6 & - \\
2013 & 9.7 & 4.8 \\
2014 & 7.9 & 4.4 \\
\hline
\end{tabular}
\end{table}

The match is not perfect though, mainly because the HDUST model is not
actually fitted to the data but just compared with it. Such a fit is beyond the scope of this paper and will be the subject of a forthcoming work (Faes et al., in prep.).

With these two models, one with a naked star and one with a gas disk,
we provide the confirmation that our images indeed exhibit a disk very
close to the star in 2014, and no disk in 2011 and 2012.

\section{Comparison between images and model} \label{sec:imgmodel}

For each set of observations, we also reconstructed an image of both the RvZ and HDUST models, as if they had been observed with PIONIER. For that, we
produced synthetic data with the same UV coverage and the same error
bars as the observations, but with values of visibilities and closure
phases replaced by those directly extracted from the RvZ and HDUST models.
The same procedure as in Sect.~\ref{sec:imaging} was applied to these
synthetic data to get reconstructed images of the model, giving us a
common basis for the comparison with the images.

We used a photosphere model that does not include
any polar wind nor any companion star, even if both could be present
in our data. 
The polar wind signature may be seen in 2011 and 2012 at the smallest
frequencies from the systematic offset seen in the residuals in
Fig.~\ref{fig:residual}. 
However, in the polar direction, we only have long baselines that are less
sensitive to an extended polar wind. The companion star signature
would produce an increased scatter along all the frequencies, as it is
relatively far away from the main star and it contributes to only a
few percent of the total H-band flux. Studying  these
aspects in detail is beyond the scope of this work.

We convolved the final images with a beam of size equal to the
effective resolution of the observations, i.e., 1.3\,mas. We normalized
the dynamic range to 1 and we centered both images before
subtracting the image of the model to the image based on the real
data. 
Such a procedure is biased if one wants to compute accurate excess
fluxes, but we lack precise H-band photometry inside and outside the
line-emission period of Achernar, which would enable us to set the
total flux of each component. Therefore, this method provides us
only a lower value of the excess flux.

The result is shown in Fig.~\ref{fig:imagesSous2} in which we present the subtraction between the images reconstructed from the observed data and the images reconstructed from the artificial data generated from the RvZ and HDUST models. For the same reason as in Section \ref{sec:hdust}, we do not show the subtraction of the HDUST images to the 2011 and 2012 images.

The subtracted images show clearly a variation of the shape of the object
during the period from 2011 to 2014. These images show no clear structure around the star in
2011 and 2012, but in 2014 there is an excess of flux on
the order of $5\%$ locally that is non-negligible for the instrument
sensibility of $\sim 1\%$. Using the HDUST model image improved the situation, confirming that this equatorial excess flux could indeed be explained by the gas disk.

This disk structure is tilted by
$18.5^\circ\pm6^\circ$ from the equatorial direction of the star. We do not know at this point if this misalignment is
significant or not, given the amount of processing we used to retrieve
this disk image.
This misalignment could be related to disk variability in Achernar, as
found by \cite{2007ApJ...671L..49C}. Indeed, blobs of matter are
expelled from the photosphere of Achernar and form rings of matter
around the star that can be detected by means of rapid polarization
angle variations from 30$^{\circ}$ to 36$^{\circ}$.

\section{Discussion and conclusions} \label{sec:discussion}

In this paper we have presented H-band images of the Be star Achernar at
different epochs corresponding to normal B and Be phases. 
These images were obtained from image reconstructed techniques applied
to VLTI/PIONIER observations taken at three different epochs (2011, 2012, and 2014). 
While the images do not show any disk (within a $\sim\pm1\%$
level) in 2011-2012 \citep[see][]{2014AandA...569A..10D}, the image
from 2014 data shows that a disk was present mainly around the
equatorial region of Achernar.
This is in line with the Achernar outburst detected in
the beginning of 2013. Follow-up spectroscopic and polarimetric observations
showed a growing disk throughout 2013 and 2014
\citep{2015IAUS..307..261F}.
\begin{figure}
\centering
\includegraphics[width=.45\textwidth]{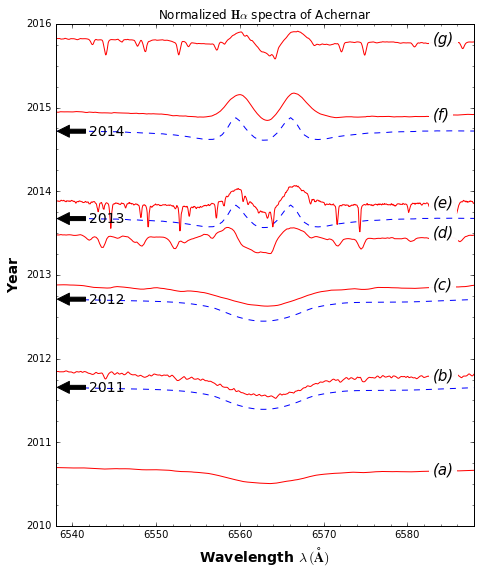}
\caption{Normalized H$\alpha$ line profiles of Achernar taken at
  different epochs between the end of 2011 and the end of
  2015\protect\footnotemark. The arrows indicate the PIONIER
  observations corresponding to Table \ref{tab:LogData}. The dashed lines
  correspond to the model H$\alpha$ line profile computed (1) with
  CHARRON (see Sect.~\ref{sec:charron}) for the photospheric RvZ model
  (epochs 2011 and 2012) and (2) with HDUST (see
  Sect.~\ref{sec:hdust}) for the photosphere and disk model (epochs 2013
  and 2014).}
\label{fig:spectra}
\end{figure}
Fig.~\ref{fig:spectra} shows several of these H$\alpha$ spectra of
Achernar taken at different epochs covering the PIONIER
observations. Details of each  spectrum are given in the caption of
the figure. H$\alpha$ is in absorption before end of 2012, so that the
2011 and 2012 PIONIER data were recorded when Achernar was essentially
in a normal B phase \citep[a thorough discussion on this phase
pre-2013 is given by][]{2014AandA...569A..10D}. On the other hand, as
shown in Fig.~\ref{fig:spectra}, H$\alpha$ is in clear emission from
mid-2013 to at least the  end of 2015, meaning that the 2013 and 2014
PIONIER data were recorded with Achernar in an emission-line
phase (Be phase).
\footnotetext{Origin of the spectra and observers:
\vspace{-0.1cm}
\begin{description}
\item[(a)] BESS 2010-09-03 (observer: Romeo), 
\item[(b)] BESS 2011-10-18 (observer: Heathcote), 
\item[(c)] OPD-ECASS-Brazil 2012-11-20 (observer: Moser Faes), 
\item[(d)] Brazil 2013-07-02 (observers: Marcon \& Napole{\~a}o), 
\item[(e)] OPD-MUSICOS-Brazil 2013-11-13 (observer: Moser Faes), 
\item[(f)] BESS 2014-11-28 (observer: Powles), 
\item[(g)] BESS 2015-10-24 (observer: Luckas).
\end{description}}

From the 2014 reconstructed image compared to the 2011 and 2012
images, we estimated that the near-IR contribution of the disk extends
to $\sim 1.7-2.3$ equatorial radii, independently from any
modeling. This result is in good agreement with predictions from
theoretical Be-disk models, providing direct support to them.

Although the question of the physical mechanism triggering the
ejection of material from the star followed by disk formation is still
unanswered, the results from this work provide direct information about
the near-IR size and flux of a newly formed Be disk, which will
contribute to future works on the disk formation and dissipation
processes in Be stars and massive stars in general.

\begin{acknowledgements}
We thank A.~Chelli for fruitful discussions and for his helpful
comments and suggestions. We thank the JMMC for all the preparation
and data interpretation tools made available to the community. We
thank E.~Thiebaut for providing the \texttt{MiRA} software to the community. This work
has made use of the BeSS database, operated at LESIA, Observatoire de
Meudon, France\footnote{\protect\url{http://basebe.obspm.fr}}; we
thank all observers who kindly provided the H$\alpha$ spectra used in
this work. A.C.C. acknowledges support from CNPq (grant 307594/2015-7)
and FAPESP (grant 2015/17967-7). Last but not least, we would like to
thank the anonymous referee for his helpful and pertinent comments,
always greatly appreciated.
\end{acknowledgements}

\vspace{-1cm}
\bibliographystyle{aa} 
\bibliography{biblio} 

\begin{thebibliography}{23}
\expandafter\ifx\csname natexlab\endcsname\relax\def\natexlab#1{#1}\fi

\bibitem[{{Carciofi}(2011)}]{2011IAUS..272..325C}
{Carciofi}, A.~C. 2011, in IAU Symposium, Vol. 272, Active OB Stars: Structure,
  Evolution, Mass Loss, and Critical Limits, ed. C.~{Neiner}, G.~{Wade},
  G.~{Meynet}, \& G.~{Peters}, 325--336

\bibitem[{{Carciofi} \& {Bjorkman}(2006)}]{Carciofi2006_v639p1081-1094}
{Carciofi}, A.~C. \& {Bjorkman}, J.~E. 2006, \apj, 639, 1081

\bibitem[{{Carciofi} {et~al.}(2007){Carciofi}, {Magalh{\~a}es}, {Leister},
  {Bjorkman}, \& {Levenhagen}}]{2007ApJ...671L..49C}
{Carciofi}, A.~C., {Magalh{\~a}es}, A.~M., {Leister}, N.~V., {Bjorkman}, J.~E.,
  \& {Levenhagen}, R.~S. 2007, \apjl, 671, 49

\bibitem[{{Dalla Vedova}(2016)}]{DallaVedova2015_PhD}
{Dalla Vedova}, G. 2016, PhD thesis, Laboratoire Lagrange, Observatoire de la
  Côte d'Azur, Universit{\' e} Nice Sophia Antipolis (France)

\bibitem[{{Domiciano de Souza} {et~al.}(2012{\natexlab{a}}){Domiciano de
  Souza}, {Hadjara}, {Vakili}, {Bendjoya}, {Millour}, {Abe}, {Carciofi},
  {Faes}, {Kervella}, {Lagarde}, {Marconi}, {Monin}, {Niccolini}, {Petrov}, \&
  {Weigelt}}]{Domiciano-de-Souza2012_v545pA130}
{Domiciano de Souza}, A., {Hadjara}, M., {Vakili}, F., {et~al.}
  2012{\natexlab{a}}, \aap, 545, A130

\bibitem[{{Domiciano de Souza} {et~al.}(2014){Domiciano de Souza}, {Kervella},
  {Moser Faes}, {Dalla Vedova}, {M{\'e}rand}, {Le Bouquin}, {Espinosa Lara},
  {Rieutord}, {Bendjoya}, {Carciofi}, {Hadjara}, {Millour}, \&
  {Vakili}}]{2014AandA...569A..10D}
{Domiciano de Souza}, A., {Kervella}, P., {Moser Faes}, D., {et~al.} 2014,
  \aap, 569, A10

\bibitem[{{Domiciano de Souza} {et~al.}(2002){Domiciano de Souza}, {Vakili},
  {Jankov}, {Janot-Pacheco}, \& {Abe}}]{Domiciano-de-Souza2002_v393p345-357}
{Domiciano de Souza}, A., {Vakili}, F., {Jankov}, S., {Janot-Pacheco}, E., \&
  {Abe}, L. 2002, \aap, 393, 345

\bibitem[{{Domiciano de Souza} {et~al.}(2012{\natexlab{b}}){Domiciano de
  Souza}, {Zorec}, \& {Vakili}}]{Domiciano-de-Souza2012_vp321-324}
{Domiciano de Souza}, A., {Zorec}, J., \& {Vakili}, F. 2012{\natexlab{b}}, in
  SF2A-2012: Proceedings of the Annual meeting of the French Society of
  Astronomy and Astrophysics, ed. S.~{Boissier}, P.~{de Laverny},
  N.~{Nardetto}, R.~{Samadi}, D.~{Valls-Gabaud}, \& H.~{Wozniak}, 321--324

\bibitem[{{Faes}(2015)}]{Faes2015_PhD}
{Faes}, D.~M. 2015, PhD thesis, IAG-Universidade de Sao Paulo (Brazil), Lab.
  Lagrange-Universit{\' e} de Nice Sophia Antipolis (France)

\bibitem[{{Faes} {et~al.}(2015){Faes}, {Domiciano de Souza}, {Carciofi}, \&
  {Bendjoya}}]{2015IAUS..307..261F}
{Faes}, D.~M., {Domiciano de Souza}, A., {Carciofi}, A.~C., \& {Bendjoya}, P.
  2015, in IAU Symposium, Vol. 307, 261--266

\bibitem[{{Glindemann} {et~al.}(2003){Glindemann}, {Algomedo}, {Amestica},
  {Ballester}, {Bauvir}, {Bugue{\~n}o}, {Correia}, {Delgado}, {Delplancke},
  {Derie}, {Duhoux}, {di Folco}, {Gennai}, {Gilli}, {Giordano}, {Gitton},
  {Guisard}, {Housen}, {Huxley}, {Kervella}, {Kiekebusch}, {Koehler},
  {L{\'e}v{\^e}que}, {Longinotti}, {M{\'e}nardi}, {Morel}, {Paresce}, {Phan
  Duc}, {Richichi}, {Sch{\"o}ller}, {Tarenghi}, {Wallander}, {Wittkowski}, \&
  {Wilhelm}}]{2003ESASP.522E...5G}
{Glindemann}, A., {Algomedo}, J., {Amestica}, R., {et~al.} 2003, in ESA Special
  Publication, Vol. 522, GENIE - DARWIN Workshop - Hunting for Planets, 5

\bibitem[{{Kervella} \& {Domiciano de Souza}(2007)}]{2007AandA...474L..49K}
{Kervella}, P. \& {Domiciano de Souza}, A. 2007, \aap, 474, 49

\bibitem[{{Kervella} {et~al.}(2008){Kervella}, {Domiciano de Souza}, \&
  {Bendjoya}}]{2008AandA...484L..13K}
{Kervella}, P., {Domiciano de Souza}, A., \& {Bendjoya}, P. 2008, \aap, 484, 13

\bibitem[{{Kluska} {et~al.}(2014){Kluska}, {Malbet}, {Berger}, {Baron},
  {Lazareff}, {Le Bouquin}, {Monnier}, {Soulez}, \&
  {Thi{\'e}baut}}]{2014AandA...564A..80K}
{Kluska}, J., {Malbet}, F., {Berger}, J.-P., {et~al.} 2014, \aap, 564, A80

\bibitem[{{Le Bouquin} {et~al.}(2011){Le Bouquin}, {Berger}, {Lazareff},
  {Zins}, {Haguenauer}, {Jocou}, {Kern}, {Millan-Gabet}, {Traub}, {Absil},
  {Augereau}, {Benisty}, {Blind}, {Bonfils}, {Bourget}, {Delboulbe},
  {Feautrier}, {Germain}, {Gitton}, {Gillier}, {Kiekebusch}, {Kluska},
  {Knudstrup}, {Labeye}, {Lizon}, {Monin}, {Magnard}, {Malbet}, {Maurel},
  {M{\'e}nard}, {Micallef}, {Michaud}, {Montagnier}, {Morel}, {Moulin},
  {Perraut}, {Popovic}, {Rabou}, {Rochat}, {Rojas}, {Roussel}, {Roux},
  {Stadler}, {Stefl}, {Tatulli}, \& {Ventura}}]{2011AandA...535A..67L}
{Le Bouquin}, J.-B., {Berger}, J.-P., {Lazareff}, B., {et~al.} 2011, \aap, 535,
  A67

\bibitem[{{Millour} {et~al.}(2012){Millour}, {Vannier}, \&
  {Meilland}}]{2012SPIE.8445E..1BM}
{Millour}, F.~A., {Vannier}, M., \& {Meilland}, A. 2012, in SPIE Conf., Vol.
  8445

\bibitem[{{Rivinius} {et~al.}(2013){Rivinius}, {Carciofi}, \&
  {Martayan}}]{Rivinius2013_v21p69}
{Rivinius}, T., {Carciofi}, A.~C., \& {Martayan}, C. 2013, \aapr, 21, 69

\bibitem[{{Thi{\'e}baut}(2008)}]{2008SPIE.7013E..43T}
{Thi{\'e}baut}, E. 2008, in SPIE Conf., Vol. 7013

\bibitem[{van Altena(2012)}]{van2012astrometry}
van Altena, W.~F. 2012, Astrometry for Astrophysics: Methods, Models, and
  Applications (Cambridge University Press)

\bibitem[{{van Leeuwen}(2007)}]{2007AandA...474..653V}
{van Leeuwen}, F. 2007, \aap, 474, 653

\bibitem[{{Vieira} {et~al.}(2015){Vieira}, {Carciofi}, \&
  {Bjorkman}}]{Vieira2015_v454p2107-2119}
{Vieira}, R.~G., {Carciofi}, A.~C., \& {Bjorkman}, J.~E. 2015, \mnras, 454,
  2107

\bibitem[{{Vinicius} {et~al.}(2006){Vinicius}, {Zorec}, {Leister}, \&
  {Levenhagen}}]{2006AandA...446..643V}
{Vinicius}, M.~M.~F., {Zorec}, J., {Leister}, N.~V., \& {Levenhagen}, R.~S.
  2006, \aap, 446, 643

\bibitem[{{von Zeipel}(1924)}]{1924MNRAS..84..684V}
{von Zeipel}, H. 1924, \mnras, 84, 684

\end{thebibliography}

\end{document}